\documentclass[sigconf]{acmart}
\AtBeginDocument{%
  \providecommand\BibTeX{{%
    \normalfont B\kern-0.5em{\scshape i\kern-0.25em b}\kern-0.8em\TeX}}}

\setcopyright{acmcopyright}
\copyrightyear{2022}
\acmYear{2022}
\acmDOI{XXXXXXX.XXXXXXX}

\acmConference[ISS'22]{Make sure to enter the correct
  conference title from your rights confirmation emai}{November 20 - 23, 2022}{Wellington, New Zealand}
\acmBooktitle{In Proceedings of ISS 2022: ACM Interactive Surfaces and Spaces, November 20 - 23, 2022, Wellington, New Zealand. ACM, New York, NY, USA} 
\acmPrice{15.00}
\acmISBN{978-1-4503-XXXX-X/18/06}

\begin{document}

\title{GesPlayer: Using Augmented Gestures to Empower Video Players}


\author{Xiang Li}
\affiliation{%
  \institution{University of Cambridge}
  \city{Cambridge}
  \country{United Kingdom}
}
\email{xl529@cam.ac.uk}

\author{Yuzheng Chen}
\affiliation{%
  \institution{Xi'an Jiaotong-Liverpool University}
  \city{Suzhou}
  \country{China}
}
\email{yuzheng.chen18@student.xjtlu.edu.cn}

\author{Xiaohang Tang}
\affiliation{%
  \institution{University of Liverpool}
  \city{Liverpool}
  \country{United Kingdom}
}
\email{sgxtang4@liverpool.ac.uk}

\renewcommand{\shortauthors}{Li, et al.}


\begin{abstract}
In this paper, we introduce GesPlayer, a gesture-based empowered video player that explores how users can experience their hands as an interface through gestures. We provide three semantic gestures based on the camera of a computer or other smart device to detect and adjust the progress of video playback, volume, and screen brightness, respectively. Our goal is to enable users to control video playback simply by their gestures in the air, without the need to use a mouse or keyboard, especially when it is not convenient to do so. Ultimately, we hope to expand our understanding of gesture-based interaction by understanding the inclusiveness of designing the hand as an interactive interface, and further broaden the state of semantic gestures in an interactive environment through computational interaction methods.
\end{abstract}

\ccsdesc[500]{Human-centered computing~Ubiquitous and mobile computing systems and tools}
\ccsdesc[500]{Human-centered computing~Gestural input}

\keywords{semantic gestures, augmented reality, gesture input, video player}

\begin{teaserfigure}
  \includegraphics[width=\textwidth]{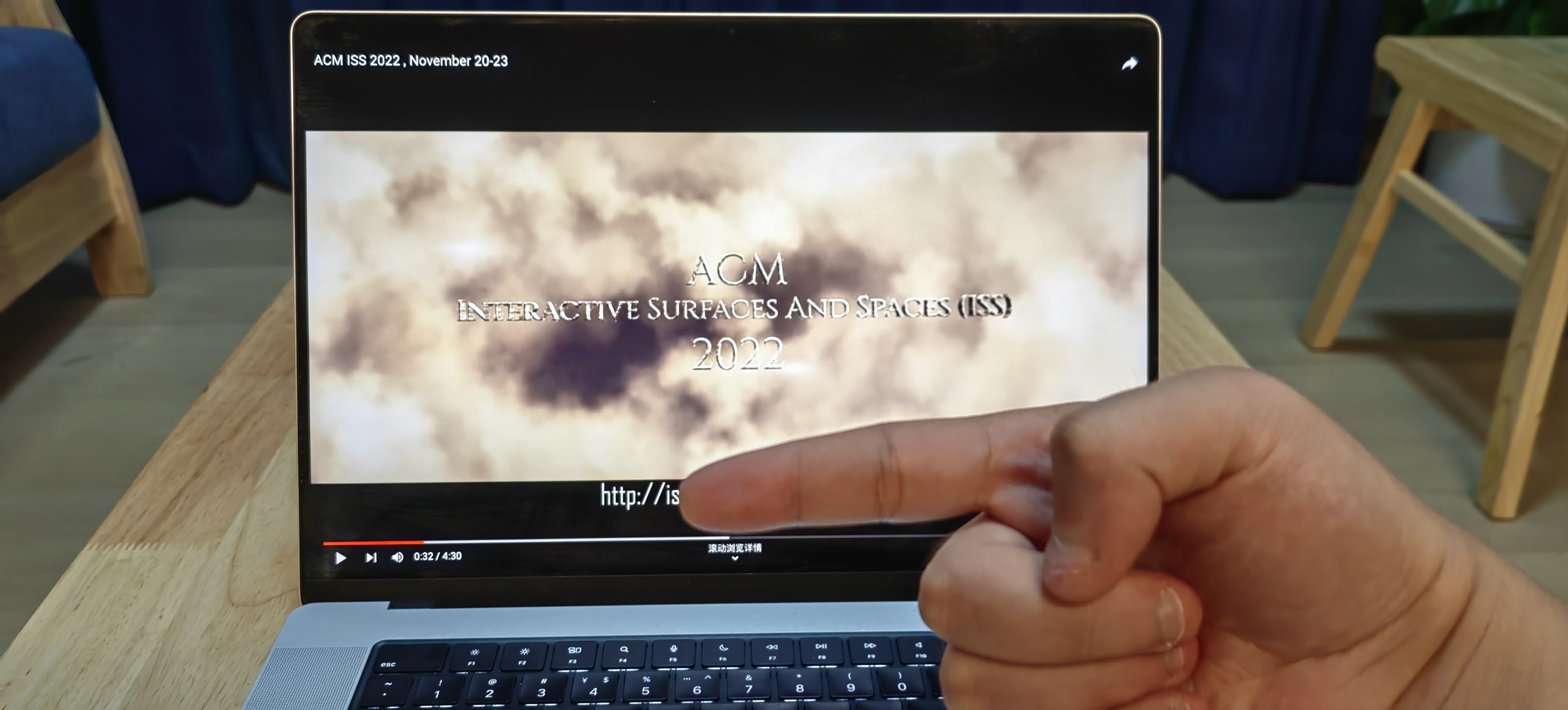}
  \caption{Example: Using GesPlayer to control the video player.}
  \Description{In the middle of the picture is a promo for ACM ISS 2022, with a hand at the bottom right, index finger extended and other fingers together.}
  \label{fig:teaser}
\end{teaserfigure}

\maketitle

\section{Introduction}
Hand-based interaction is one of the most commonly used interaction methods in human-computer interaction (HCI) and intelligent interactive systems \cite{malik2002hand}. Currently, most of the AR and VR head-mounted displays also implement the function of mid-air gesture interaction for more natural interaction. In addition, semantic gestures are being explored to understand the underlying gestural behavior during user interactions in order to design and propose more intuitive ways of interaction. To fully investigate the importance of semantic gestures, Adam Kendon has summarized how the gestured component contributes to the meaning or expression of the utterance \cite{kendon2004gesture}.

However, gestures are more than just an input to the process of interaction, but also instructions based on changes and activities of the body (especially the fingers). In the process of moving our fingers, we often overlook the expressive interface that is part of the body and the vehicle of gesture: the hand itself. Thus most gestures are designed to please the movement itself, which does improve the experience of interaction to some extent, but ignores the independence of the hand as an interface.

In this paper, we present GesPlayer, an augmented gesture-based video player which explores how users can also experience their hands as an interface with their gestures. We articulate the technical setup of GesPlayer, its design, and present preliminary findings, and give our discussions in the form of two themes: (a) how the hand as an interface expresses the interaction of gestures, and (b) the state of hand interaction. Ultimately, we hope to expand our understanding of gesture-based interaction by understanding the inclusion of designing hands as an interactive interface and further broaden the state of semantic gestures in interactive environments via the computational interaction method.

\section{Related Work}

\subsection{Mid-air Gestures}
Koutsabasis and Vogiatzidakis point out that mid-air interactions are characterized by (a) touchless interactions, (b) real-time sensor tracking of (parts of) the user's body, and (c) body movements, postures, and gestures need to be recognized and matched to specific user intentions, goals, and commands \cite{koutsabasis2019empirical}. 

Most previous research on gesture-based interaction has been based on the use of one or more RGB cameras \cite{biswas2011gesture}. For example, Dani et al. proposed a low-cost approach using only a single monocular RGB camera \cite{dani2018mid}. Similarly, Jain et al \cite{jain2019gestarlite} proposed a low-cost framework to manipulate objects in mid-air using only one RGB camera. In summary, with the recent advances in low-cost depth cameras and RGB cameras, many algorithms and techniques (see \cite{chen2017survey}) have been developed to enable gesture recognition for mid-air interaction.

\subsection{Semantic Gestures}
Gestures are a vehicle for conveying semantic information and an input for the external environment to understand the gestures. Thus, the research of semantic gestures focuses on semantics itself rather than on gesture design. 

In semantic gestures, there is a term, Utterance, which refers to any combination of actions that counts to others as an attempt by the actor to "provide" some kind of information \cite{kendon2004gesture}. Drawing on Goffman's formulation \cite{goffman1981forms}, he notes that although whenever people are present together with each other they cannot avoid providing each other with information about their intentions and involvement, their status as social beings, and their own personal character, and thus can be said to be "giving out" information, people also engage in actions that are considered to be explicitly aimed at providing information.

\subsection{Experiencing Body in HCI}
Game research in HCI has always been interesting in the human body. Mueller et al. suggest that the field of game research is evolving from using keyboards to play digital content to using the body to play digital content, moving toward a future where we experience the body as a digital game \cite{mueller2018experiencing}. To guide designers interested in supporting players to experience their bodies as games, Mueller et al. present two phenomenological perspectives on the human body (Körper and Leib) \cite{mueller2018experiencing}. Mueller et al. argue that before new sensors entered the realm of game design, users primarily used mice and keyboards, joysticks and gamepads to play computer games. With the advancement of sensors such as Kinect, users are beginning to use their bodies to experience digital content (Körper). Ultimately, the vision presented by Mueller et al. is that we are able to experience our bodies as digital games (Körper \& Leib). In addition, another paper by Mueller et al. investigates the possibility of "limited control of the body" as an interesting design resource for body play systems and proposes four strategies for limited control of the body: Exploration, Reflection, Learning, and Embracement \cite{floyd2021limited}.

\section{GesPlayer}

\subsection{Technical Setup}

We used Unity 2021.2.13f1 for developing our system. The detection algorithm and the parameters collection were empowered by the MediaPipe \footnote{\href{https://google.github.io/mediapipe/getting_started/python}{MediaPipe} is a framework for building pipelines to perform inference over arbitrary sensory data. With MediaPipe, a perception pipeline can be built as a graph of modular components, including model inference, media processing algorithms and data transformations, etc. \cite{lugaresi2019mediapipe}} APIs and its Unity Plugin \footnote{\href{https://github.com/homuler/MediapipeUnityPlugin}{https://github.com/homuler/MediapipeUnityPlugin}}. 

\subsection{Design}
In GesPlayer, we introduced three different states of gestures: "trigger gestures", "baseline gestures", and "interaction gestures": (a) a trigger gesture is defined in GesPlayer as a prerequisite for starting and triggering a judgment. This safeguards the balance between gestural interactions and normal bodily movements, and reduces misdetection for our system; (b) the baseline gesture is defined as a benchmark to calibrate some behaviors that require relatively precise operations: for example, adjusting the progress bar to control the progress of the currently playing video. When traditionally using the mouse or keyboard to control the progress of the video playback, we would see a reference at the bottom of the video (i.e., the progress bar), but the lack of an intuitive control for mid-air interactions would also make it difficult to move the video or perform other relatively precise operations. Therefore, the purpose of baseline gestures is to support truly interaction gestures to achieve more precise results (i.e., the act of dragging and dropping); and (c) the definition of interaction gestures is the same as in previous studies. That is, the action in effect performs the user's real command. For example, we manipulate objects in real life \cite{patibanda2021actuating}, use gestures for text input \cite{lu2020exploration}, translate the objects in dynamic tasks \cite{xu2020exploring}, experience the world in VR and navigate using the senses of touch, motion restriction, and proprioception \cite{vm2021touch}, fight the virus in VR boxing games \cite{xu2020virusboxing}, or use our bodies for gesture detection in virtual environments \cite{li2021vrcaptcha}.

\subsubsection{Progress of Video Playback}
\begin{figure}[t]
  \centering
  \includegraphics[width=0.45\textwidth,scale=0.5]{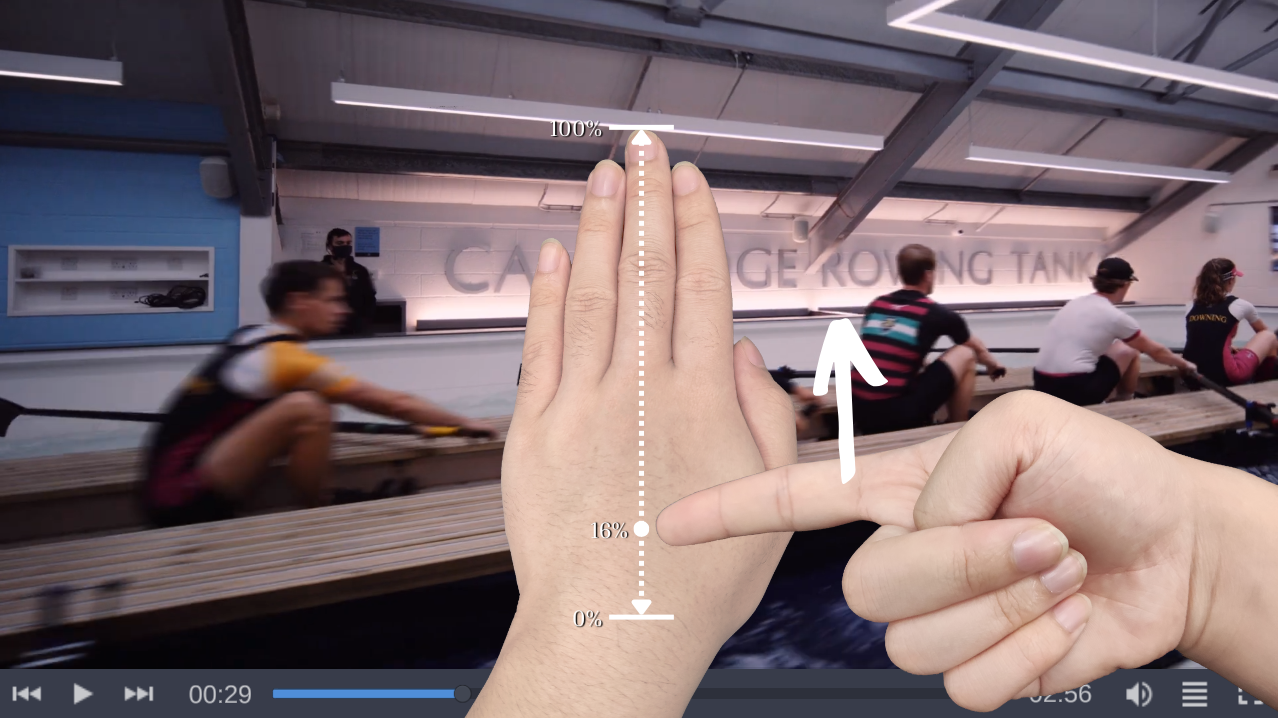}
\includegraphics[width=0.45\textwidth,scale=0.5]{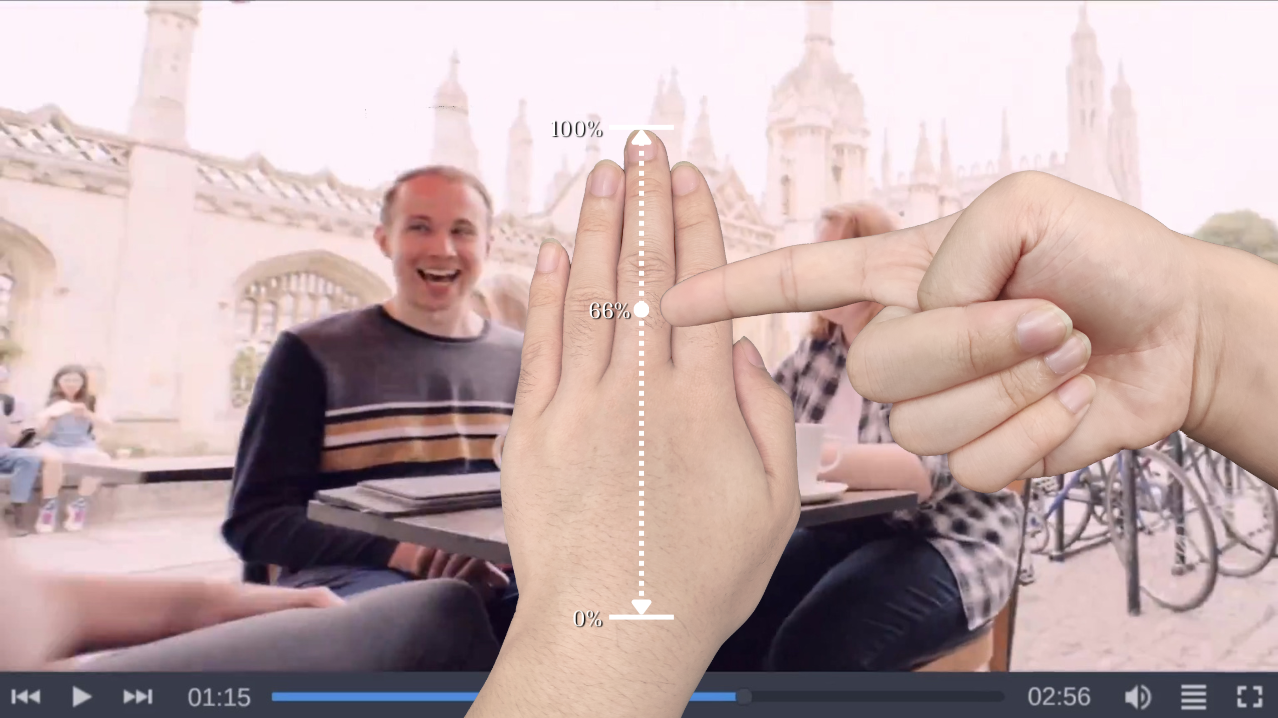}
  \caption{In the video playback control, we used the left hand as the progress bar and the right hand as the pointer. Once the left hand stays in front of the camera, the directed line segments between the wrist and the middle finger tip would stand for the progress bar of the video.}
  \Description{In the video playback control, we used the left hand as the progress bar and the right hand as the pointer. Once the left hand stays in front of the camera, the directed line segments between the wrist and the middle finger tip would stand for the progress bar of the video.}
  \label{fig:VideoPlayback}
\end{figure}

In the video playback control, we used the left hand as the progress bar and the right hand as the pointer. Once the left hand stays in front of the camera, the directed line segments between the wrist and the middle finger tip would stand for the progress bar of the video. The wrist and the middle finger tip stand for 0\% and 100\% of the progress of video playback. Once the progress bar is set up, use the right-hand index fingertip as the pointer to adjust the video progress by hiding the right-hand thumb and touching the directed line segments on the left hand physically. Touching for starting the adjustment and untouching to finish the adjustment. Real-time progress bar feedback will be shown on the screen while the adjustment is activated (see Figure \ref{fig:VideoPlayback}).

\subsubsection{Video Volume}

\begin{figure}[h]
  \centering
  \includegraphics[width=0.45\textwidth]{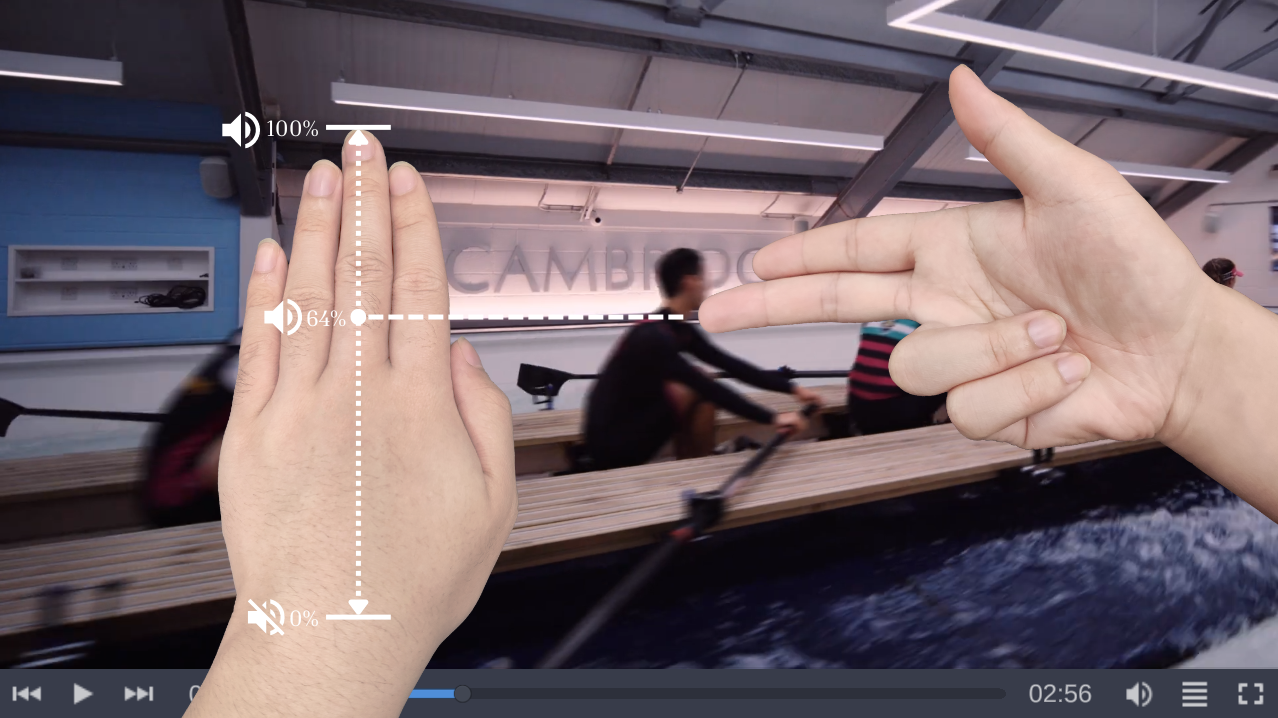}
  \includegraphics[width=0.45\textwidth]{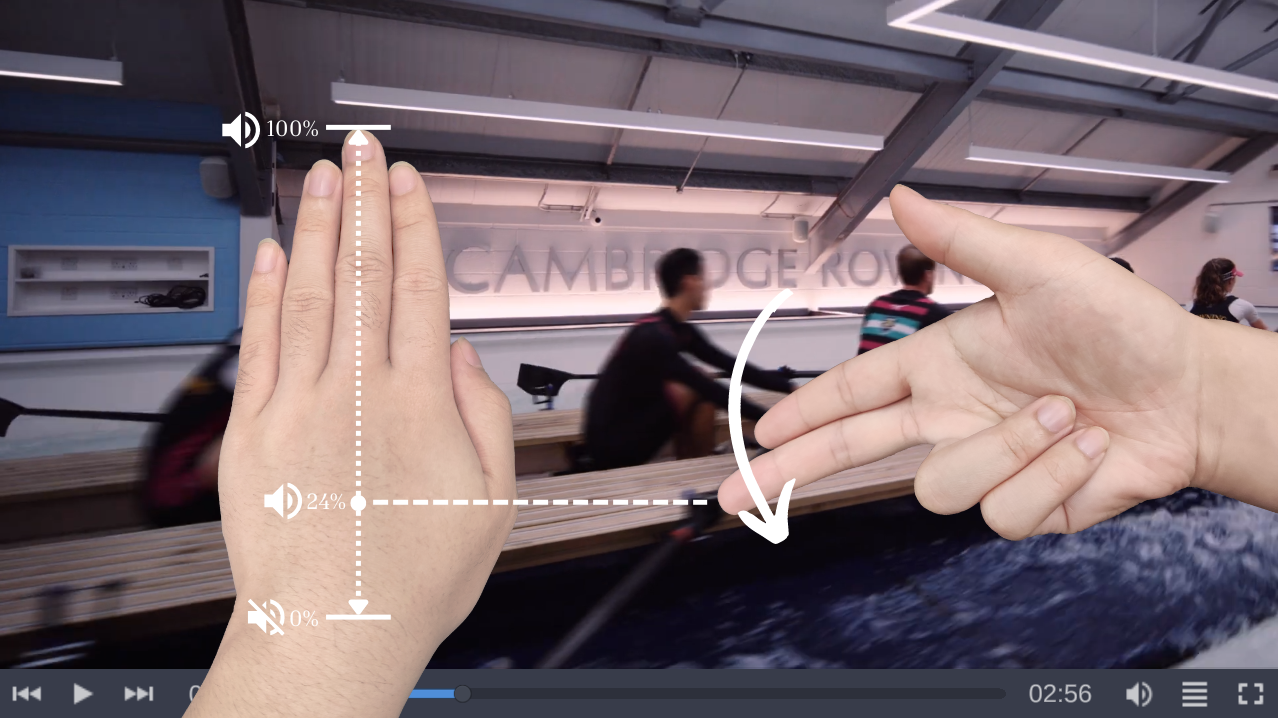}
  \caption{In the video volume control, we keep the left hand as the volume bar, however, use the "thumb", "index" and "middle" finger gestures on the right hand and use the middle fingertip as the pointer to adjust the volume.}
  \Description{In the video volume control, we keep the left hand as the volume bar, however, use the "thumb", "index" and "middle" finger gestures on the right hand and use the middle fingertip as the pointer to adjust the volume.}
  \label{fig:VideoVolume}
\end{figure}

In the video volume control, we keep the left hand as the volume bar, however, we use the "thumb", "index" and "middle" finger gestures on the right hand and use the middle fingertip as the pointer to adjust the volume. The left hand should also keep distance from the right hand in this case, and the volume value also depends on the closest point from the right-hand index fingertip to the directed line segments between the left hand's wrist and the middle fingertip. The volume would be also changed in real-time while adjusting to the feedback (see Figure \ref{fig:VideoVolume}).

\subsubsection{Screen Brightness}

\begin{figure}[h]
  \centering
  \includegraphics[width=0.45\textwidth]{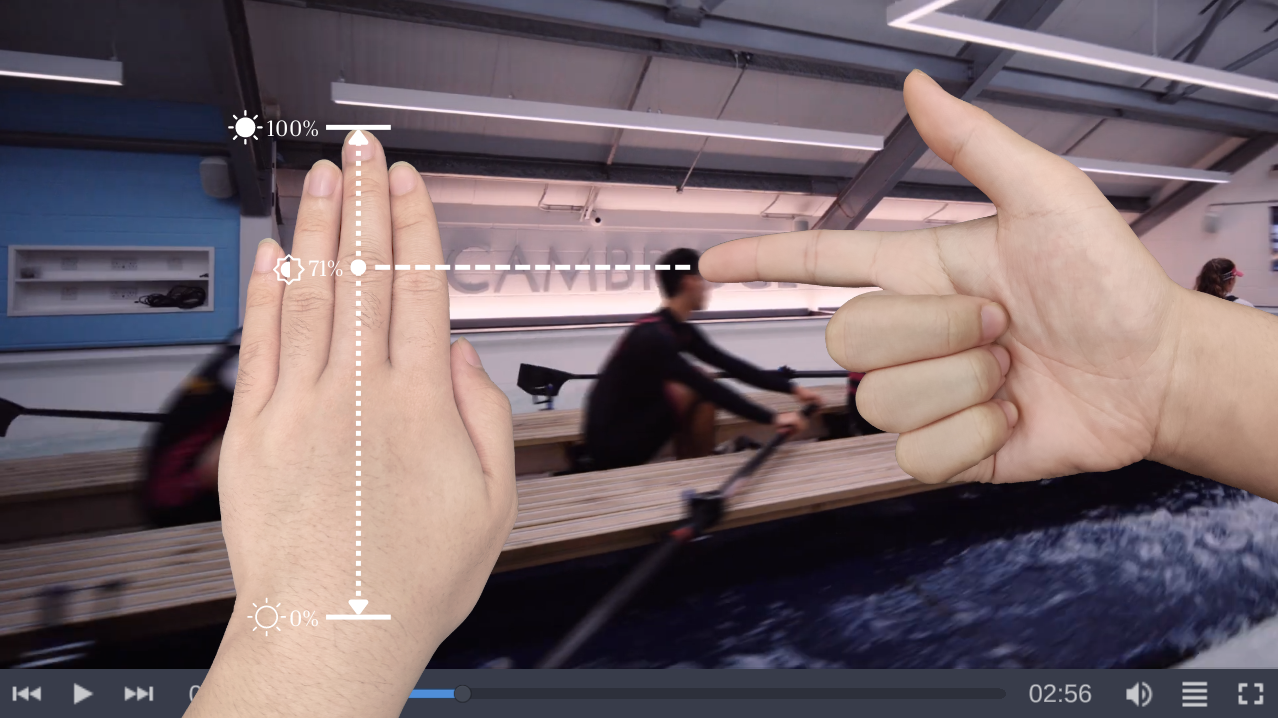}
  \includegraphics[width=0.45\textwidth]{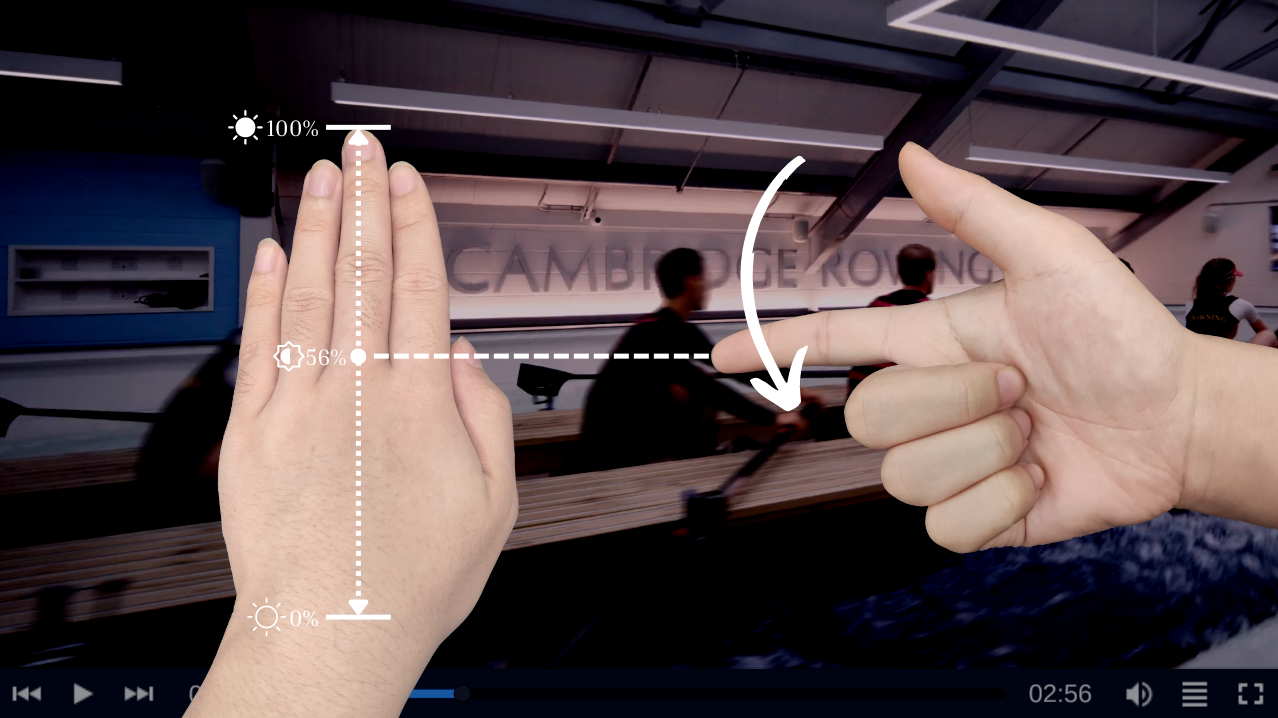}
  \caption{In the screen brightness control, we used the left hand as the brightness bar and the right hand as the pointer. Similarly, once the left hand appears in front of the screen, the directed line segments between the wrist and the middle finger tip would be recognized as the brightness bar and the wrist and the middle finger tip stands for 0\% and 100\% of the brightness of the screen.}
  \Description{In the screen brightness control, we used the left hand as the brightness bar and the right hand as the pointer. Similarly, once the left hand appears in front of the screen, the directed line segments between the wrist and the middle finger tip would be recognized as the brightness bar and the wrist and the middle finger tip stands for 0\% and 100\% of the brightness of the screen.}
  \label{fig:ScreenBrightness}
\end{figure}

Similar to the video progress control, in the screen brightness control, we used the left hand as the brightness bar and the right hand as the pointer. Similarly, once the left hand appears in front of the screen, the directed line segments between the wrist and the middle finger tip would be recognized as the brightness bar and the wrist and the middle finger tip stands for 0\% and 100\% of the brightness of the screen. Once the brightness bar is set up, make the right hand in the gesture of only revealing the thumb and index finger. Keep distance between left hand and right hand and use the right-hand index fingertip as the pointer to adjust the brightness. The closest point from the right-hand index fingertip to the directed line segments between the left hand's wrist and the middle finger tip would be the value of the screen brightness. The brightness would be changed in real-time while adjusting to the feedback (see Figure \ref{fig:ScreenBrightness}).

\section{Preliminary Findings and Discussions}

\subsection{Hands As an Interface to Express Gestures}
HCI research has recognized human skin as a promising interface for interacting with intelligent computing devices \cite{steimle2017skin}. Its use as an interface helps to overcome the limited surface space of today's wearable devices and allows for input to multiple smart devices. Most existing work treats the skin as a hypothetical flat surface, with the principles and models for designing interactions shifting from existing touch-based devices to the skin. In addition, current skin interactions typically allow only touch gestures or taps in several different distinct locations, thus greatly limiting the possible interactions to expressive interactions with a wide range of user interfaces and applications. However, in GesPlayer, instead of directly bringing the two hands into physical contact, our design allows one hand to assist the hand that needs to provide the gesture to better express the semantic information. We believe that a visual benchmark will significantly improve the accuracy of dynamic tasks that require gestural movement.

\subsection{The State of Gestural Interaction}
According to the concept of "Computational Interaction" \cite{oulasvirta2018computational}, the essence of interaction is the direct interplay of information expressed by behaviors and actions in different states. In previous studies, the state of gestural input/output (I/O) has been defined as only one step in the overall interaction process. In fact, the description of the state should be more clearly defined: in GesPlayer, we introduced "trigger gestures", "baseline gestures", and "interaction gestures", which represented the "start" state, the "benchmark" state, and the "input/output" state. For the user, this series of state changes is the real behavioral analysis of interaction, which could be helpful for researchers to understand the underlying behavioral semantic information.

\section{Conclusion and Future Work}
In this paper, we introduce a gesture-based empowered video player, GesPlayer, which explores how users can experience their hands as an interface through gestures. We provide three semantic gestures based on the camera of a computer or other smart device to detect and adjust the progress of video playback, volume, and screen brightness, respectively. We articulate the technical setup of GesPlayer, its design, and present preliminary findings, and give our discussions in the form of two themes: (a) how the hand as an interface expresses the interaction of gestures, and (b) the state of hand interaction. Our goal is to enable users to control video playback simply by their gestures in the air, without the need to use a mouse or keyboard, especially when it is not convenient to do so. Ultimately, we hope to expand our understanding of gesture-based interaction by understanding the inclusiveness of designing the hand as an interactive interface, and further broaden the state of semantic gestures in an interactive environment through computational interaction methods.

In the future, we plan to conduct a user study to evaluate our GesPlayer in three areas: (a) usability and efficiency, (b) advantages and potential disadvantages of GesPlayer compared to gesture interaction without a baseline state, and (c) how semantic gestures can support a better mastery of our control over the video player.

\begin{acks}
Xiang Li is supported by the China Scholarship Council (CSC) International Cambridge Scholarship (No. 202208320092).
\end{acks}

\bibliographystyle{ACM-Reference-Format}
\bibliography{gesplayer}

\end{document}